\documentclass[%
 reprint,
superscriptaddress,
 amsmath,amssymb,
 aps,
pra,
]{revtex4-2}
\renewcommand{\selectlanguage}[1]{}
\usepackage[caption=false]{subfig}
\usepackage{float}
\usepackage{graphicx}
\usepackage{subfig}  
\usepackage{dcolumn}
\usepackage{bm}
\usepackage{tikz} 
\usetikzlibrary{quantikz2}
\usepackage{braket} 

\begin{document}

\title{Improved amplitude amplification strategies for the quantum simulation of classical transport problems}

\author{Alessandro Andrea Zecchi}
\affiliation{%
MOX – Department of Mathematics \\
Politecnico di Milano \\
Piazza L. da Vinci, 32, 20133 Milano, Italy
}%
\author{Claudio Sanavio}
\affiliation{
Center for Life Nano-Neuroscience at la Sapienza \\
Fondazione Istituto Italiano di Tecnologia \\
Viale Regina Elena 291, Roma, 00161, Italy
}%
\author{Simona Perotto}%
\affiliation{%
MOX – Department of Mathematics \\
Politecnico di Milano \\
Piazza L. da Vinci, 32, 20133 Milano, Italy
}%
\author{{Sauro Succi}}
\affiliation{
Center for Life Nano-Neuroscience at la Sapienza \\
Fondazione Istituto Italiano di Tecnologia \\
Viale Regina Elena 291, Roma, 00161, Italy
}%

\date{\today}

\begin{abstract}
  The quantum simulation of classical fluids often involves the use of probabilistic algorithms that encode the result of the dynamics in the form of  the amplitude of the selected quantum state. In most cases, however, the amplitude probability is too low to allow an efficient use of these algorithms, thereby hindering the practical viability of the quantum simulation. The oblivious amplitude amplification algorithm is often presented as a solution to this problem, but to no avail for most classical problems, since its applicability is limited to unitary dynamics. In this paper, we show analytically that oblivious amplitude amplification when applied to non-unitary dynamics leads to a distortion of the quantum state and to an accompanying error in the quantum update. We provide an analytical upper bound of such error as a function of the degree of non-unitarity of the dynamics and we test it against a quantum simulation of an advection-diffusion-reaction equation, a transport problem of major relevance in science and engineering. Finally, we also propose an amplification strategy that helps mitigate the distortion error, while still securing an enhanced success probability. 
\end{abstract}

\maketitle

\section{Introduction}

Since its conceptual inception \cite{feynman1982simulating}, potential applications of
quantum computing have been identified across a broad variety of problems
involving quantum systems in science and engineering \cite{Steijl_2024,di_meglio_quantum_2024,alexeev_quantum-centric_2024}. Recently, there has been an increased interest in using quantum algorithms to simulate
classical physics problems and in particular those described by non-linear partial differential equations, fluid dynamics being an outstanding example in point
\cite{succi2023quantum,zhao_quantum_2022,steijl_quantum_2019,sanavio_quantum_2024}. 
The main advantage of using quantum computing comes from the linear
superposition of states and the resulting inherent parallelism of quantum operations.
Indeed, by exploiting superposition, a $q$ qubits quantum register can store
$2^q$ complex amplitudes, each of which can be updated in parallel
thanks to the non-local nature of quantum mechanics. However, developing quantum computational fluid dynamics solutions remains a particularly challenging task due to the inherently linear, unitary, and non-dissipative nature of quantum operations, which conflicts with the characteristics of fluid flow \cite{sanavio2024three}. 

Several strategies have been devised in the recent years to deal with  nonlinearity and dissipation \cite{berry2015hamiltonian,low2019hamiltonian,childs2012hamiltonian,sanavio_quantum_2024}.
In particular, Carleman embedding as applied to the lattice Boltzmann formulation of fluid flows
has proven particularly attractive in terms of convergence of the classical procedure replacing
the original nonlinear problem with a truncated sequence of linear ones \cite{fluids7010024}.
Unfortunately, the corresponding quantum circuit is plagued by an
exponential depth of the corresponding quantum circuit once projected onto the native Pauli
quantum gates.
In a recent work, taken here as a study case, the dynamics of a one-dimensional
advection-diffusion-reaction (ADR) system has been  encoded using the Carleman embedding
\cite{sanavio2024explicit} and shown to reduce the exponential complexity
to a quadratic one by resorting to block-encoding of sparse matrices, a
technique borrowed from Quantum Control and Quantum Signal Processing
to embed a non-unitary matrix into a unitary operator
\cite{caneva2009optimal,khodjasteh2010arbitrarily,low_optimal_2017}. 

The block encoding technique has been proposed as a candidate to address the non-unitarity issue
thereby offering an appealing route to advance the development of quantum algorithms for fluid flows
\cite{low2019hamiltonian,gilyen_quantum_2019,li_potential_2025}. However, block-encoding comes at the cost of a low success probability
of the corresponding probabilistic update.
In principle, this can be obviated by the oblivious amplitude amplification (OAA) algorithm, were if not for the fact that
OAA only works for unitary operations. More precisely, the application of OAA
to non-unitary matrices leads to a distortion of the resulting state.
To the best of our knowledge, this distortion effect has been only pointed out ~\cite{berry2015simulating} but not analysed in quantitative terms.

In this paper we analyse explicitly the introduced distortion by studying the effect of the OAA algorithm onto a block encoding based circuit for a one-dimensional ADR system.  In more detail, we propose a way to mitigate the error
while ensuring an increase of the success probability.
Our strategy relies on an enhanced amplitude amplification algorithm based on
a computationally cheap approximation of the system's evolution to
perform the standard reflection about the initial state. The paper is organized as follows.
In section \ref{sec1}, we present the basics of the block-encoding procedure
by highlighting the associated stochastic nature.
Amplitude amplification strategies are reviewed in section \ref{sec2}, emphasizing the need for an
oblivious amplification algorithm, which is then analysed in section \ref{sec3}.
The effect of the non-unitarity on OAA is further discussed and a new model for
the error is introduced in section \ref{new}, where possible error-mitigating strategies are also
analysed. Additionally, a novel method based on an approximate form of standard amplitude amplification is proposed, providing a closer proxy to the exact solution. In section \ref{num} we present numerical results
and provide evidence of the superior performance of the proposed method versus the standard OAA algorithm.

\section{Block encoding} \label{sec1}

Classical systems, such as those representing transport problems, are affected by dissipative and non linear effects that make their dynamics highly non unitary. This characteristic poses a serious challenge to the quantum simulation of classical dynamics, as quantum computers can operate only in a unitary fashion. 
As a possible remedy, one can implement the dynamics of interest by defining the non-unitary operation as a building block of a larger unitary matrix acting on an extended Hilbert space. This technique, generally called \textit{block-encoding}, was developed in the context of developing Hamiltonian simulation algorithms ~\cite{low2019hamiltonian}. 
More specifically, given the non-unitary matrix $A\in \mathbb{C}^{2^n\times 2^n}$ where $n$ is the number of qubits of the target register, we define the  unitary matrix  $U \in \mathbb{C}^{2^{n+m}\times 2^{n+m}}$ that block-encodes $A$ with $m$ ancillary qubits as
\begin{equation}
    U=
    \begin{bmatrix}
        A/\alpha & * \\
        * & *
    \end{bmatrix},
    \label{eq:1}
\end{equation}

\noindent where the elements $*$ guarantee the unitarity of $U$ while the scaling factor $\alpha$ ensures that the singular values of any submatrix of $U$ are bounded by 1 \cite{camps2024explicit}. 
It is straightforward to show that when $U$ is applied to the state $\ket{0}\ket{\psi}$, we get the following result
\begin{equation}
    U \ket{0}\ket{\psi} = c_0\ket{0}A\ket{\psi}+\sum_{i=1}^{2^m-1} c_i \ket{i}\ket{*}
\end{equation}
where the coefficients $c_i$ depend on the specific values of the matrix blocks indicated with the symbol $*$.
This means that the non-unitary matrix $A$ is applied conditionally to the vector $\ket{\psi}$ when all ancillary qubits are measured in the state $\ket{0}$. Thus, the success probability is simply given by $p(0)=c_0^2\Vert A \ket{\psi} \Vert^2$.

Amplitude amplification algorithms have been proposed to overcome the probabilistic nature of block encoding with the aim of increasing the probability of measuring the ancillary state $\ket{0}$~\cite{chakraborty2018power}.

\section{Amplitude amplification strategies} \label{sec2}
In many quantum computing applications the solution to a problem is often encoded in a specific state of a larger wave function describing the overall register. To increase the probability of measuring a desired state, various amplitude amplification algorithms \cite{brassard1998quantum, brassard2002quantum, ambainis2010variable} have been developed based on the framework of Grover's search algorithm, which offers a quadratic speed-up over the best-known classical search methods \cite{grover1996fast}.

A fundamental component of these types of algorithm is the repeated application of a specific operator, which is designed to rotate the full state vector in the two-dimensional space spanned by a target and an orthogonal state. In more detail, the so-called \textit{Grover operator} is defined as 
\begin{equation}
    G=UR_sU^{\dagger}R_t
    \label{eq:G}
\end{equation}
where $U$ indicates the state preparation operator, $U^\dagger$ its inverse while $R_t$ and $R_s$ are reflection operators about the target and initial states. However, a rotation that goes beyond this target state reduces the probability of success, which is  referred to as the \textit{souffl\'e problem}, since iterating too few times \textit{undercooks} the result, but iterating too many times \textit{overcooks} it. As a consequence, there is an optimal number of Grover iterations that need to be performed to reach the maximum amplitude, which could still be lower than 1. Furthermore, in order to apply Grover's iteration the correct number of times, one has to know the original amplitude of the target state, an information that is not always accessible to the user. A different strategy is offered by fixed-point algorithms, which do not suffer from the over rotation problem and monotonically increase the amplitude of the target state \cite{grover2005fixed,mizel2009critically, yoder2014fixed}.

We remark that the exact knowledge of the initial state of the system is a strict requirement for the correct application of standard and fixed point amplitude amplification algorithms. However, in many cases of interest the initial state cannot be known in advance, and an alternative \textit{blind} approach has to be used. 
An oblivious amplitude amplification (OAA) algorithm that does not require any knowledge of the initial state preparation of the quantum working register has been proposed in the context of Hamiltonian simulation techniques \cite{berry2014exponential}. More recently, it has been developed a new fixed point OAA algorithm \cite{yan2022fixed} that monotonically increase the success probability by repeatedly applying a Grover operator based on the Linear Combination of Unitaries (LCU) method \cite{childs2012hamiltonian}. This algorithm combines OAA with a damping mechanism initially proposed by Mizel in \cite{mizel2009critically}. 

\section{Oblivious amplitude amplification for non unitary matrices} \label{sec3}

 The standard OAA algorithm uses $m$ ancillary qubits and employs two unitary matrices $U$ and $V$, defined on $n+m$ and $n$ qubits, respectively. In particular $U$ acts, for an angle $\theta \in (0,\frac{\pi}{2})$, on a generic state $\ket{\psi}$ in the following way
\begin{equation}
    U\ket{0}\ket{\psi}=\sin\theta\ket{0}V\ket{\psi}+\cos\theta\ket{\Phi^\perp},
    \label{eq:3}
\end{equation}

\noindent where $\ket{\Phi^\perp}$ is a state that depends on $\ket{\psi}$ and is orthogonal to any state characterized by the $\ket{0}$ state of the $m$ ancillary qubits. By defining the operator 
\begin{equation}
    S=-URU^{\dagger}R,
    \label{eq:S}
\end{equation} 

\noindent where $R=2P-\mathbb{I}$ is the reflection operator with respect to the state $\ket{0}$ of the ancilla qubits and $P=\ket{0}\bra{0}\otimes\mathbb{I}$ is the corresponding projector, the following relation, defined for any $l\in \mathbb{Z}$, has been proven \cite{berry2014exponential}:
\begin{equation}
    S^lU\ket{0}\ket{\psi}=\sin((2l+1)\theta)\ket{0}V\ket{\psi}+\cos((2l+1)\theta)\ket{\Phi^\perp}.
\end{equation}
After applying the $S$ operator $l$ times, the success probability of measuring the target state $\ket{0}V\ket{\psi}$ changes from the initial value $|\sin\theta|^2$ to $|\sin((2l+1)\theta)|^2$. An optimal value of $l$ can be chosen to maximize this probability. By comparing Eqs. \eqref{eq:1} and \eqref{eq:3}, we see that they are closely related, even though the matrix $A$ in ~\eqref{eq:1} is not unitary. 

In \cite{berry2015simulating}, the authors studied the effect of non-unitarity when the matrix $V$ is close to a unitary matrix. This result has been extended to a robust version of OAA in \cite{berry2015hamiltonian}. Summarizing their findings, setting $\sin\theta=1/s$ and provided that $|s-2|=O(\delta)$, it has been proved the following relation

\begin{equation}
    PSU\ket{0}\ket{\psi}=\ket{0}(\frac{3}{s}V-\frac{4}{s^3}VV^\dagger V)\ket{\psi},
    \label{eq:modstate}
\end{equation}
which shows that an error is introduced at each step of OAA since $V^\dagger V$ is not equal to the identity matrix $\mathbb{I}$.  

Using the spectral matrix norm $\Vert \cdot \Vert $ , defined for a generic matrix $M$ as $\Vert M \Vert = \sqrt{\lambda_{\max}(M^\dagger M})$,  we introduce the non-unitarity parameter
\begin{equation}
        \eta = \Vert V^\dagger V-\mathbb{I}\Vert.
        \label{eq:eta}
\end{equation}
If $V$ is $\delta$-close to a unitary matrix, with $\eta=O(\delta)$, then the error can be bounded as follows
\begin{equation}
    \Vert PS^lUP-\ket{0}\bra{0}\otimes V \Vert = O(\delta).
    \label{eq:err}
\end{equation}

\section{A new perspective on oblivious amplitude amplification for non-unitary matrices} \label{new}
The OAA algorithm is based on the 2D Subspace Lemma \cite{berry2014exponential} which states that, defined the state $\ket{\Psi}=\ket{0}\ket{\psi}$ and the effect of $U$ on this state as in Eq.~\eqref{eq:3}, it is then possible to define an orthogonal state, that also satisfies the property $P\ket{\Psi^\perp}=0$ and s.t.
\begin{equation}
    U\ket{\Psi^{\perp}}=\cos\theta\ket{0}V\ket{\psi}-\sin\theta\ket{\Phi^\perp}.
    \label{eq:perp}
\end{equation}
The orthogonality of the two states can be proved by taking the inner product of Eq.~\eqref{eq:3} and Eq.~\eqref{eq:perp} resulting in $\braket{\Psi|\Psi^\perp}=0$.  
However, for a non-unitary matrix $V$, the orthogonality condition of the state defined in Eq.~\eqref{eq:perp} is no longer satisfied, thus leading to an error after the application of the operator $S$. Nevertheless, an orthogonal state still exists and it is characterized by the following expression:
\begin{equation}
     U\ket{\Psi^{\perp}_{\text{true}}}=\cos\theta\ket{0}(V^\dagger)^{-1}\ket{\psi}-\sin\theta\ket{\Phi^\perp}
     \label{eq:perp2}.
\end{equation}
Note that this is equivalent to the previous definition whenever $V$ is unitary.  The proof can be obtained by simply taking the inner product with Eq.~\eqref{eq:3} and  using the fact that $V^\dagger (V^\dagger)^{-1}=\mathbb{I}$. Equation \eqref{eq:perp2} suggests that there might be a way to cancel out the error by using different reflection operators $R$ in \eqref{eq:3} or by replacing $U^\dagger$ with a different \textit{inverse} operation, as discussed in Sec.~\ref{sec3}.

Using the updated definition \eqref{eq:perp2} for the orthogonal state and following the same procedure of Ref.~\cite{berry2014exponential}, we are able to obtain an approximate solution (see Appendix~\ref{appendix}) after one iteration of the OAA algorithm, that is more general than the expression in ~\eqref{eq:modstate} since $\sin\theta$ is not constrained to values close to 1/2. This  yields: 
\begin{equation}
\begin{split}
    SU\ket{0}\ket{\psi}\simeq\sin3\theta\ket{\Phi}+\cos3\theta\ket{\Phi^{\perp}}\\
    -2\sin\theta \cos^2\theta\ket{\epsilon}-2\sin\theta \cos^2\theta URU^\dagger\ket {\epsilon},
\end{split}
\label{eq:appr}
\end{equation}
where the state  $\ket{\epsilon}=\ket{\Phi}-\ket{\Phi_\epsilon}$ is a combination of the exact $\ket{\Phi}=\ket{0}V\ket{\psi}$ with the approximate $\ket{\Phi_\epsilon}=\ket{0}(V^\dagger)^{-1}\ket{\psi}$ state. It is also possible to provide the following expression for $\ket{\epsilon}$, which highlights the effect of the non-unitarity
\begin{equation}
    \ket{\epsilon}=\ket{0}\otimes(V-(V^\dagger)^{-1})\ket{\psi}=\ket{0}\otimes (V^\dagger)^{-1}(V^\dagger V-\mathbb{I})\ket{\psi}.
    \label{eq:eps-def}
\end{equation}
With reference to expression \eqref{eq:appr}, we find the typical amplitude amplification terms involving functions $\sin(3\theta)$ and $\cos(3\theta)$ along some perturbations that depend on $\ket{\epsilon}$, which vanish as $V$ approaches unitarity. 

We now introduce a simple linear model to characterize the error introduced by OAA. 
First, we define the exact normalized target state $\ket{\beta}=V\ket{\psi}/\Vert V \ket{\psi}\Vert$ and its orthogonal counterpart $\ket{\beta^\perp}$. It is possible to realize that the error-introducing terms in \eqref{eq:appr} are characterized by the state $\ket{\epsilon}$, which depends on the application of the operator $V^\dagger V-\mathbb{I}$ to $\ket{\psi}$ ( cf. Eq.~\eqref{eq:eps-def}). \\
Therefore, by taking the projection along the $\ket{0}$ state of the ancillary qubits, without loss of generality, we obtain:
\begin{equation}
    \ket{\omega}=PSU\ket{0}\ket{\psi} = \frac{\ket{\beta}+c \eta \ket{\beta^\perp}}{\sqrt{1+(c \eta)^2}}
    \label{eq:proj}
\end{equation}
where $c$ is a parameter that depends on $\ket{\psi}$, on the initial angle $\theta$ of expression \eqref{eq:appr} and on the matrix $V$.  Roughly speaking $c$ is a parameter which accounts for the amplitude of the component orthogonal to the exact solution after one step of OAA. Note that the error shows a linear dependence on $\eta$, as shown by the numerical verificaton in Sec.\ref{num}.

\subsection{The Euclidean distance}
We now use Eq.~\eqref{eq:proj} to retrieve the Euclidean distance between the exact solution $\ket{\beta}$ and the resulting state after one OAA iteration $\ket{\omega}$.
Since $\braket{\beta|\beta} =1$ and $\braket{\beta|\beta^\perp}=0$ we conclude that the Euclidean distance is simply given by:
\begin{equation}
    D(\beta, \omega)=\sqrt{2(1-|\braket{ \beta| \omega}|)}=\sqrt{2\bigg{(}1-\frac{1}{\sqrt{1+(c \eta)^2}}\bigg{)}}.
    \label{eq:dist-func}
\end{equation}
This value is limited between $0$, the ideal error-free case achieved when the non-unitarity parameter goes to 0, and $\sqrt{2}$, corresponding to the maximum possible error.  
Furthermore, for small values of the non-unitarity parameter $\eta$, the error is bounded by a linear function of $\eta$ as it is possible to verify in Figure \ref{fig:first-func}, retrieving an equivalent version of the error bound of \eqref{eq:err}.
We found an upper bound for the Euclidean distance $D$ by setting in \eqref{eq:appr}  
\begin{equation}
\begin{aligned}
    \ket{\epsilon}\simeq\eta \ket{0}\ket{\beta ^\perp}&, &  URU^\dagger\ket{\epsilon}\simeq\ket {\epsilon} .
\end{aligned}
\label{eq:assum}
\end{equation}
With these assumptions, the Euclidean distance is maximised to 

\begin{equation}
    D\leq D_{max}=\sqrt{2(1-\frac{1}{\sqrt{1+(\frac{4\sin\theta\cos^2\theta}{\sin3\theta} \eta)^2}})}.
    \label{eq:dmax}
\end{equation}

\subsection{The fidelity}
\begin{figure}[htbp]  
    \centering
    \subfloat[]{\includegraphics[width=0.92\linewidth]{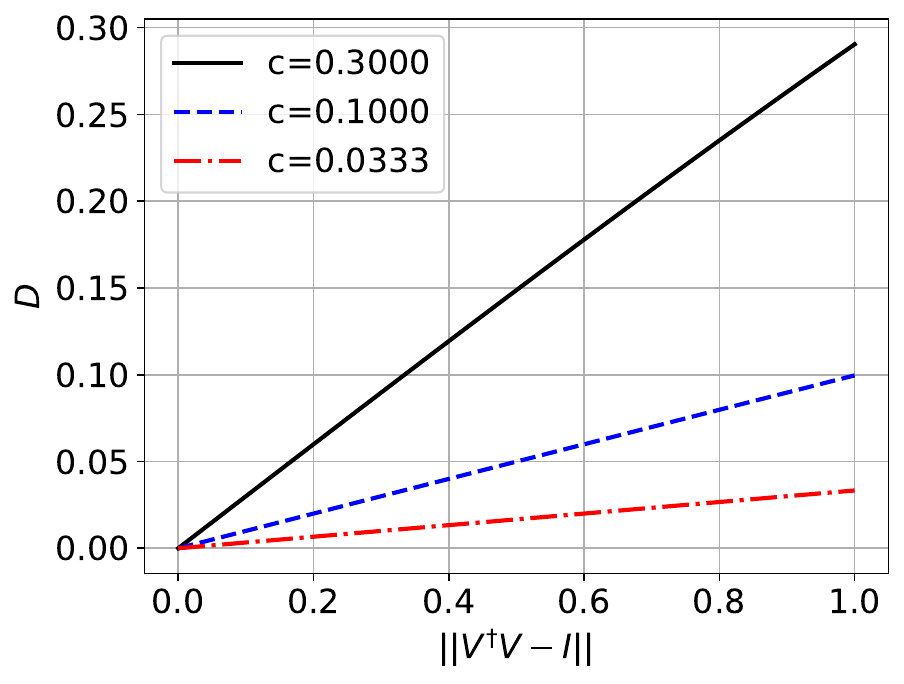}
    \label{fig:first-func}}
    \hfill
    \subfloat[]{\includegraphics[width=0.92\linewidth]{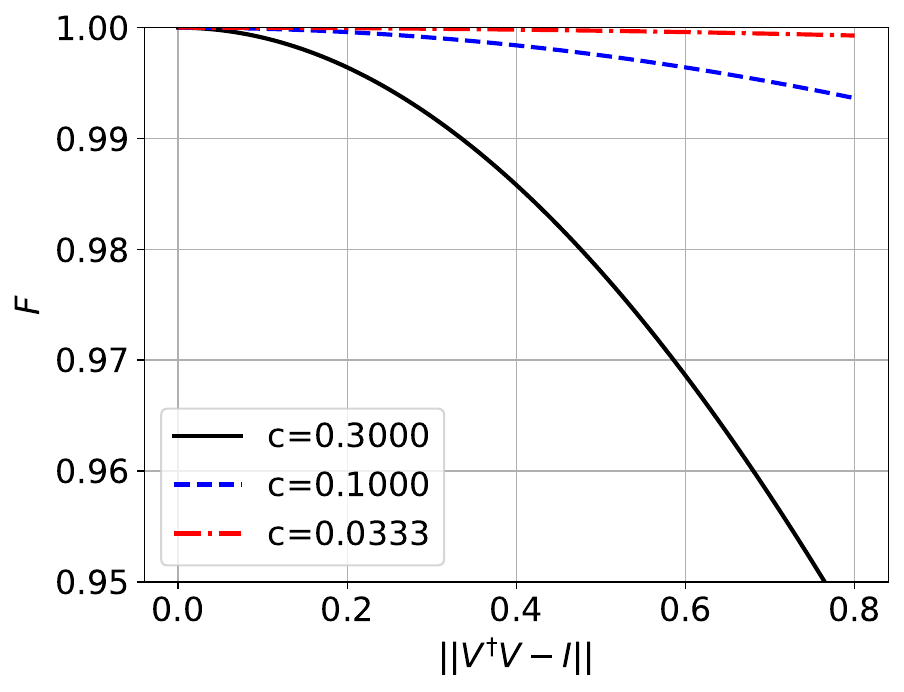}
    \label{fig:func}}
    \caption{Trend of the euclidean distance $D$ (a) and fidelity $F$ (b) as a function of the non-unitarity parameter for 3 representative values of $c$. The ideal case would be an error-free situation with $D=0$ and $F=1$.}
    \label{fig:two_functions}
\end{figure}

To complete the analysis of the distortion error, we next present a model for the fidelity of the quantum state after OAA when dealing with a non-unitary matrix $V$. After one step of the amplification process by using \eqref{eq:proj}, we can write the fidelity between $\ket{\omega}$ and $\ket{\beta}$ as:
\begin{eqnarray}
    F(\beta,\omega) &=&  |\braket{\beta | \omega}|^2 = \left| \frac{\braket{\beta|\beta }+c \eta \braket{\beta|\beta^\perp }}{\sqrt{1+(c \eta)^2}}\right|^2\\
    &=& \frac{1}{1+(c \eta)^2}.
    \label{eq:fid-func}
\end{eqnarray}
The value $1$ for $F$ corresponds to an exact solution without the introduction of any distortion. If we plot $F$ over the non unitarity parameter $\eta$ (Figure \ref{fig:func}), we obtain a decreasing Bell-shaped curve, whose width depends on the value of the parameter $c$. As the non-unitarity goes to 0, the fidelity gets closer to $1$, recovering the ideal case.
Under the same assumptions in \eqref{eq:assum}, we can provide a lower bound for the fidelity, yielding,
\begin{equation}
    F\geq F_{min}=\frac{1}{1+({\frac{4\sin\theta\cos^2\theta}{\sin3\theta}} \eta)^2}.
    \label{eq:fmin}
\end{equation}

\subsection{Error-mitigation strategies and approximate reflection amplitude amplification} \label{error-mitigation}
The distortion introduced at each iteration by the OAA algorithm when dealing with non-unitary matrices represents a strong limitation to the development of quantum algorithms aiming at efficiently implementing a non-unitary dynamics through block encoding.
The main issue of applying OAA to non-unitary matrices can be stated as follows: when the matrix $V$ is non-unitary, applying the operator $U^\dagger$ to the state $\ket{0}V\ket{\psi}$ fails to map it back to the initial state, as
\begin{equation}
    U^\dagger\ket{0}V\ket{\psi}=\sin\theta\ket{0}V^\dagger V\ket{\psi}+\sum_{i=1}^{2^m-1} c_i \ket{i} \ket{*}.
\end{equation}
In general, we are looking for a modified version of the OAA algorithm that substitutes $S$  in ~\eqref{eq:S} with the routine $\tilde S= -URU_2R$, leading to a reduced (potentially zero) error and to an increased success probability.
A possibility is to replace $U_2$ with the operator $W$ that block encodes $V^{-1}$, such that
\begin{equation}
    W\ket{0}V\ket{\psi}=\sin(\theta_W)\ket{0}V^{-1}V\ket{\psi}+\sum_{i=1}^{2^m-1} c_{i,W} \ket{i} \ket{*}.
\end{equation}
However, the numerical investigation we carried out shows that this strategy does not provide any practical advantage, whereas the efficient implementation of $W$ is not straightforward. 
At the same time, we proved that it is not possible to find a unitary operator $U_2$ that solves the error problem and simultaneously allows for the amplification of the target state.
Thus, the development of a completely error-free OAA algorithm for non-unitary matrices remains an open problem, requiring a fundamentally different framework. \\
Here, we propose an alternative approach based on an approximate method,  which we show to provide better results for our use case as it leads to a reduced error with respect to the OAA algorithm. 

Standard amplitude amplification requires a reflection about the initial state $R_s$ and 
another reflection about the target subspace $R_t$. 
However, when the target is identified by the state $\ket{0}$ in the index register (this is the case e.g. of a block-encoded operation) we can replace $R_t$ with the opposite reflection about the state $|0\rangle$ of the index register $-R$, as done in the OAA algorithm. To simplify the Grover’s algorithm even further, we propose to substitute the reflection operator $R_s$ with a reflection about an approximate state $\tilde R_s$, which is easy to compute. Thus, our modified Grover algorithm (compare with ~\eqref{eq:G}), becomes:   
\begin{eqnarray}
    \tilde G&=&U\tilde R_sU^\dagger \tilde R_t\\ \label{eq:approx_G}
    \tilde{R}_t&= & -R = - (2P-\mathbb{I})\\
    \tilde{R}_s&=&2\ket{\tilde{\psi}}\bra{\tilde{\psi}}-\mathbb{I} \label{eq:approx_R}
\end{eqnarray}
where $P$ is the projection to the $\ket{0}$ state of the ancillary qubits and $\tilde R_t=-R $ is simply the opposite of the reflection with respect to that state . This corresponds simply to flip the sign of the target state, namely a very cost-efficient and easy to implement operation. Concerning $\tilde R_s$ it coincides with a reflection operator with respect to an approximate initial solution $\ket{\tilde\psi}$. \\
It is straightforward to show that if $\Vert R_s-\tilde{R}_s \Vert=O(\delta)$, then the error between the exact solution and the state after one application of $\tilde{G}$ is such that:
\begin{equation}
    \Vert P(U\tilde{R}_sU^\dagger \tilde R_t U)P-\ket{0}\bra{0}\otimes V \Vert = O(\delta).
\end{equation}
Therefore, using a sufficiently accurate estimate of the initial state, the error introduced by the proposed algorithm is lower than the one introduced by OAA defined in  \eqref{eq:err}.

\section{Simulating the advection-diffusion-reaction dynamics with Qiskit} \label{num}

In this section we quantum-simulate an advection-diffusion-reaction dynamics \cite{sanavio2024explicit} by employing the efficient block encoding circuit of a sparse matrix \cite{camps2024explicit}. We focus on a one-dimensional domain and we discretize the time dependent advection-diffusion-reaction problem 
\begin{equation}
\begin{cases}
    \displaystyle\frac{\partial \phi }{\partial t}=D\frac{\partial ^2 \phi }{\partial x^2}-U\frac{\partial\phi}{\partial x}-a\phi   & x\in (0,L), t>0\\
    \phi(x,0)=\phi_0 & x\in (0,L)\\
    \phi(0,t)=\phi(L,t) & t>0
\end{cases}
\end{equation}
where $\phi$ is the physical quantity of interest, with uniform diffusion coefficient $D$, constant velocity $U$ and constant reaction $a$, provided with periodic boundary conditions and an initial condition $\phi_0$. Using a second order centred finite difference scheme to approximate both the derivatives on the uniform distribution of spatial nodes $x_0=0<x_1 < ... <x_{N-1}<x_{N}=L$, the problem is discretized with $N$ coordinates from $x_0$ to $x_{N-1}$:
\begin{equation}
\begin{split}
    &\dot{\phi_j} =\frac{D}{\Delta x^2}(\phi_{j-1}-2\phi_j+\phi_{j+1})-\frac{U}{2\Delta x}(\phi_{j+1}-\phi_{j-1})-a\phi_j \\
    \\
   & \forall j=0,...,N-1
\end{split}
\label{eq:ad}
\end{equation}
with $\Delta x=L/N$ the uniform grid spacing and where $\dot{\phi}_j=\displaystyle\frac{d\phi_j}{dt}=\displaystyle\frac{d\phi(x_j,t)}{dt}$. The corresponding algebraic form is given by
\begin{equation}
\dot{\bm{\phi}}(t) = M \bm{\phi}(t)
\end{equation}
with $M\in \mathbb{R}^{N\times N}$ and $\bm\phi(t)\in \mathbb{R}^N$. From now on, without loss of generality, we take $N=2^n$ where $n$ is the number of qubits used to embed the numerical problem on a quantum register. By using a forward Euler scheme to discretize the time dependence, we obtain
\begin{equation}
    \bm{\phi}(t_i+\Delta t)=(\mathbb{I}+\Delta t M)\bm{\phi}(t_i) = A \bm{\phi}(t_i)
    \label{eq:time}
\end{equation}
for each time-step $t_i$ of the temporal discretization starting from $t_0=0$. The resulting matrix $A\in\mathbb{R}^{N\times N}$ is a banded circulant matrix with the following structure:
\begin{equation}
    A =
    \begin{bmatrix}
        \lambda_0 & \lambda_1 & 0 & \dots &\lambda_2 \\ 
        \lambda_2 & \lambda_0 & \lambda_1 & \ddots & 0 \\
        0 & \ddots & \ddots & \ddots & \vdots \\ 
        \vdots & \ddots & \lambda_2 & \lambda_0 & \lambda_1 \\
        \lambda_1 & 0 & \dots & \lambda_2 & \lambda_0 \\
    \end{bmatrix}
    \label{eq:matrixA}
\end{equation}
By introducing the three dimensionless Courant numbers for diffusion, advection and reaction \cite{sanavio2024explicit, courant1928partiellen}
\begin{equation}
    \gamma_d = \frac{\Delta t D}{\Delta x ^2},\quad
    \gamma_a = \frac{\Delta t U}{\Delta x} , \quad
    \gamma_r = a \Delta t
\end{equation}
we can write the $\lambda_i$ terms in a compact form respectively as
\begin{equation}
    \lambda_0=1-2\gamma_d-\gamma_r, \quad
    \lambda_1=\gamma_d-\frac{\gamma_a}{2}, \quad
    \lambda_2=\gamma_d+\frac{\gamma_a}{2}.
\end{equation}
To guarantee the stability of the explicit finite difference scheme in time, we properly combine the CFL conditions for the Courant numbers as in ~\cite{moin_fundamentals_2010}, thus obtaining the constraints
\begin{eqnarray}
 \gamma_d \leq 1/2 ,\quad \gamma_a \leq 1, \quad \gamma_r \leq 1.
\label{eq:cfl-conditions}
\end{eqnarray}
The time-step limitation in \eqref{eq:cfl-conditions}, can be circumvented using implicit schemes or other formulations for transport phenomena \cite{diotallevi2009capillary}.
In addition, we remark that the forward Euler scheme results in a non-unitary matrix $A$ also in the case of a unitary differential operator, such as an advection problem. 

\begin{figure}[htbp]  
    \centering
    \begin{quantikz}
        \lstick{$\ket{0}$} & \qwbundle{m} & 
        \gate[wires=2][1.7cm]{U}
        \gateinput[1]{}
        \gateoutput[wires=1]{} & \meter{\ket{0}} \\
        \lstick{$\ket{\psi}$}& \qwbundle{n} & \gateinput[1]{}
        \gateoutput[wires=1]{} & & \rstick{$A\ket{\psi} / \Vert A\ket{\psi} \Vert  $} \\
    \end{quantikz}
    \caption{Schematics of the quantum circuit to block encode the matrix $A$ with $m$ ancillary qubits and $n$ working qubits. }
    \label{fig:circ}
\end{figure}
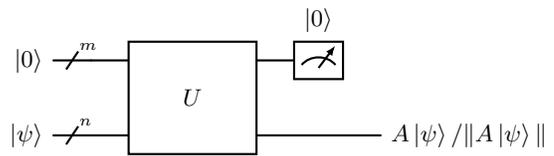

\begin{figure}[htbp]  
    \centering
    \subfloat[]{\includegraphics[width=0.92\linewidth]{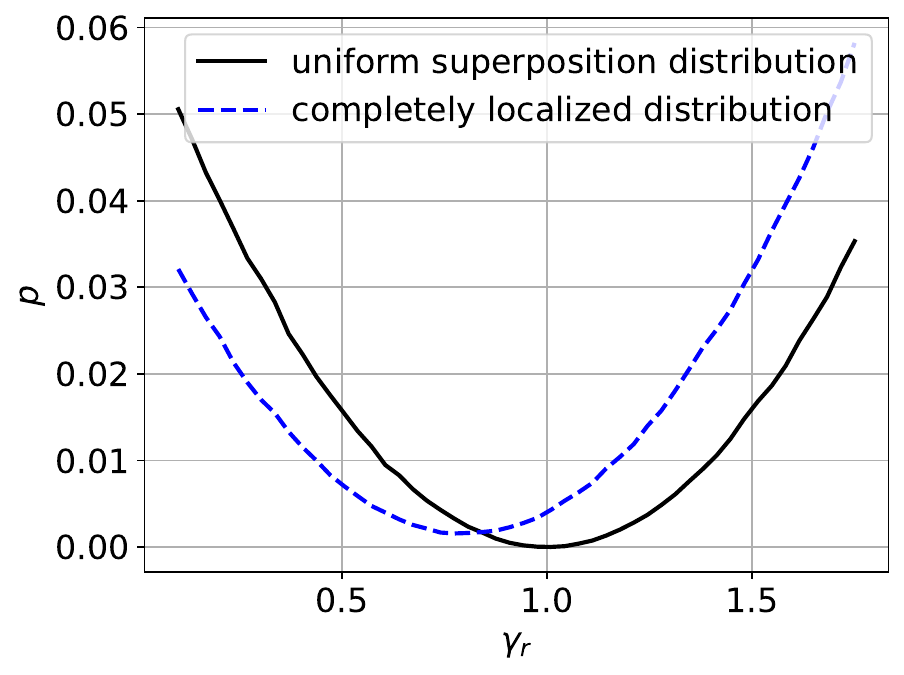}
    \label{fig:prob}}
    \hfill
    \subfloat[]{\includegraphics[width=0.92\linewidth]{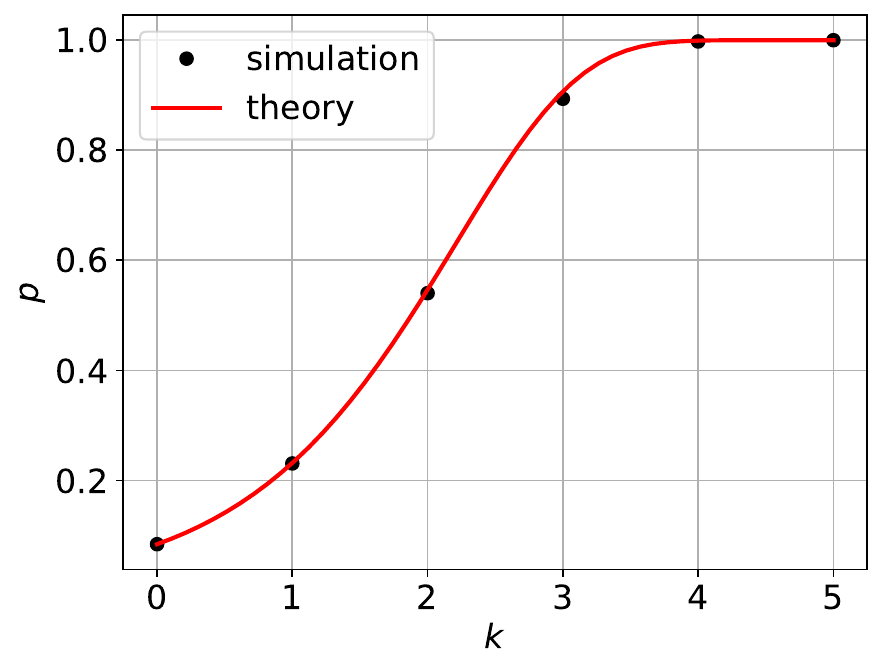}
    \label{fig:pi3_prob}}
    \caption{(a) Qiskit simulation of the success probability of implementing the operator $A$ as a function of $\gamma_r$ with $\gamma_d$ and $\gamma_a$ both set to $0.1$, using a uniform distribution (solid line) and a completely localized initial state (dashed line). (b) Qiskit simulation of the success probability (black dots) using the $\pi/3$ fixed-point algorithm as a function of the number of iterations $k$ using $\gamma_r=0.9$, $\gamma_a=\gamma_d=0.01$ with a completely localized initial state, compared to the theoretical curve (solid red line).}
    \label{fig:two_probabilities}
\end{figure}

The matrix $A$ can be block-encoded using only $m=3$ ancillary qubits, by efficiently taking advantage of its sparsity pattern. The quantum circuit to perform this operation is described in detail in \cite{sanavio2024explicit} with a specific analysis on the probabilistic nature of the final result. In particular, the block encoding scheme employed embeds the matrix $A$ with a scaling factor of $1/4$ into a larger unitary operator. As a consequence, the success probability of implementing a single time step as in \eqref{eq:time} is of the order of $1/16$, thus inhibiting the practical use of the proposed scheme to implement multiple time steps. In order to analyse the effect of different amplitude amplification strategies, we have implemented a Qiskit code~\cite{qiskit2024} that is able to generate an efficient quantum circuit to block-encode any banded circulant matrix, or symmetric $2\times2$ matrix, using the resources provided by \cite{camps2024explicit}. The developed code is currently available in \cite{zecchi_url}. A sketch of the developed circuit is shown in Figure \ref{fig:circ} where $U$ is the gate that encapsulates the block encoding of $A/4$. If the $m=3$ ancillary qubits are measured in state $\ket{0}$, then the working register has a final state-vector which is equal, up to a normalization factor, to the application of matrix $A$ to the initial state $\ket{\psi}$. Using the Qiskit platform we are able to simulate the probability of success as a function of $\gamma_r$ (Figure \ref{fig:prob}), with $\gamma_d=\gamma_a$ both set to 0.1 and $n=4$ working qubits, for both a completely localized initial state and for a uniform superposition initial state, retrieving the parabolic profile calculated analytically in \cite{sanavio2024explicit}.  In both cases the probability has a value that is too small for simulating multiple time steps.

Figure  \ref{fig:pi3_prob} confirms that by applying standard amplitude amplification which requires an exact knowledge of the initial state, no error is introduced, and the success probability can be increased. Thus, using for instance the fixed point $\pi/3$ Grover's algorithm, we are able to amplify the success probability up to a value that gets monotonically closer to 1 as the number of the algorithm iterations is increased. The simulation employs $n=3$ working qubits while setting $\gamma_r=0.9$, $\gamma_a=\gamma_d=0.01$. More precisely, if the initial success probability is $p=(1-\epsilon)$ it becomes $p=(1-\epsilon^{3^k})$ after $k$ iterations\cite{grover2005fixed}.
Nevertheless, for simulating multiple time steps, an oblivious algorithm is needed since it is not possible to know a priori $\bm{\phi}(t_i)$ for $t_i\neq0$ and to apply the appropriate exact reflection with respect to the initial state $R_s$. Hereafter, we present numerical simulations based on the OAA algorithm and on an approximate implementation of the amplitude amplification algorithm based on partial knowledge of the initial state.

\subsection{OAA for a completely localized initial state}
\begin{figure}[htbp]  
    \centering
    \begin{quantikz}
        \lstick{$\ket{0}$} & \qwbundle{m} & 
        \gate[wires=2][0.8cm]{U}  & \gate[wires=1][0.8cm]{-R} & \gate[wires=2][0.8cm]{U^\dagger}
         & \gate[wires=1][0.8cm]{R} & \gate[wires=2][0.8cm]{U} &\\
        \lstick{$\ket{\psi}$}& \qwbundle{n} & \gateinput[1]{}
        \gateoutput[wires=1]{}  & \gateinput[1]{} \gateoutput[wires=1]{} & \qw & \gateinput[1]{} \gateoutput[wires=1]{} & \qw & \\
    \end{quantikz}
    \caption{Quantum circuit for one iteration of the OAA algorithm, applied after the operator $U$ that block encodes matrix $A$.}
    \label{fig:oaa-scheme}
\end{figure}
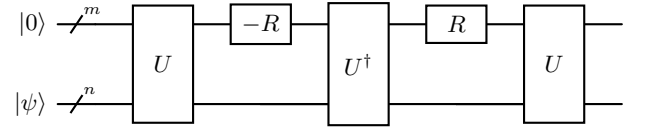

\begin{figure}[htbp]  
    \centering
    \includegraphics[width=0.92\linewidth]{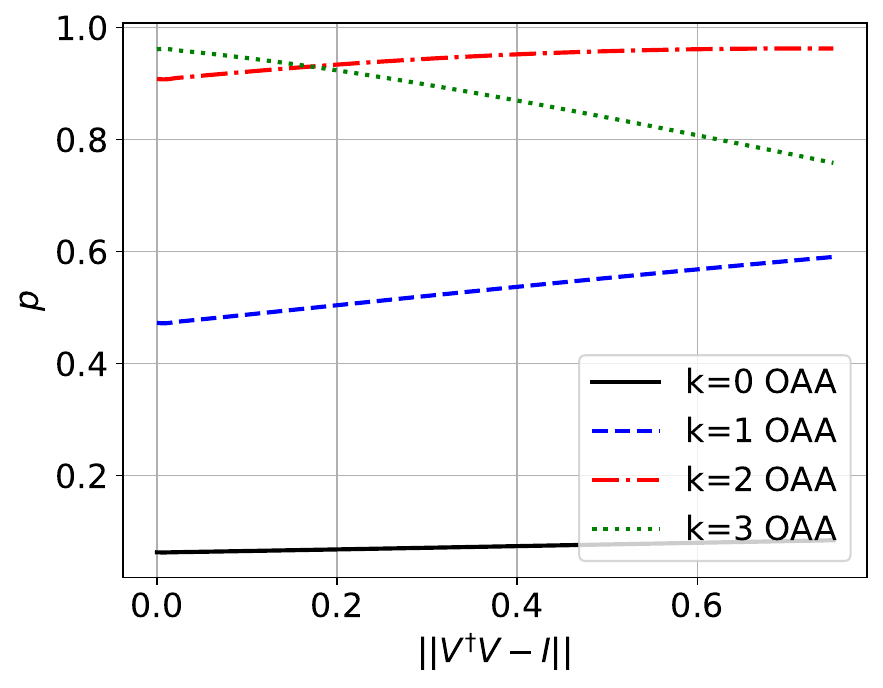}
\caption{Qiskit simulation of success probability of measuring the $\ket{0}$ state in the ancillary qubits as a function of the non-unitarity parameter $\eta=\Vert V^\dagger V - \mathbb{I}\Vert$. As expected, the optimal number of iterations is in the order of $\sqrt{8}$, therefore corresponding to $k=2$ or $3$.}
\label{fig:oaa_prob1}
\end{figure}

The following analysis is based on the possibility of changing the non-unitarity parameter $\eta$  by selecting a different time step $\Delta t$ , while preserving the value for all the other variables involved in the definition of the Courant numbers. In particular, if we set to $\gamma_d=\gamma_r=0.01$ and $\gamma_a=0.9$, the value of the non-unitarity parameter is $\eta=\Vert V^\dagger V-\mathbb{I}\Vert\simeq0.75$. Reducing the value of the time step $\Delta t$ the matrix defined in equation \eqref{eq:matrixA} approaches the identity operator and correspondingly $\eta$ goes to 0. We start with the analysis of a completely localized initial state for a working register of $n=4$ qubits. This leads to a state-vector with only one component different from zero.
If $U$ is the operator that represents the block-encoding circuit shown in Figure \ref{fig:circ}, then the OAA algorithm can be used by applying $k$ times the $S$ operator defined in  \eqref{eq:S}, interleaving $U$ and its inverse $U^\dagger$ with reflections about the state $\ket{0}$ of the three auxiliary bits (see Figure \ref{fig:oaa-scheme}).
Figure \ref{fig:oaa_prob1} shows how the probability $p=|\sin((2k+1)\theta)|^2$ is affected by changing the iteration number $k$. The optimal number of iterations $k_{opt}$ should maximize $\sin((2k+1)\theta)$, where $\theta$ depends on different factors, such as the block-encoding technique, the  Courant numbers, the choice for the initial state and the non-unitarity of the matrix. Thus, $k_{opt}\sim \mathcal{O}(\sqrt{N/M})$ where $N=2^m=8$ is the total number of states, and $M=1$ is the number of target states \cite{yan2022fixed} so that, in our case, $k_{opt}$ is either 2 or 3.  Our results show that the non-unitarity parameter has a small effect on the overall probability improvements at successive iterations.

\begin{figure}[htbp]  
    \centering
    \subfloat[]{\includegraphics[width=0.92\linewidth]{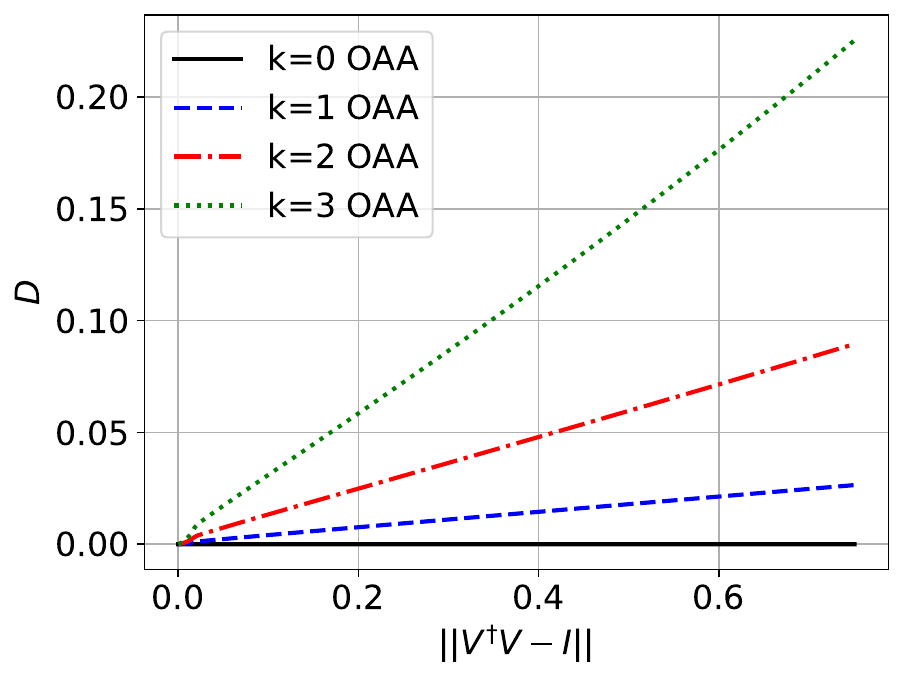}
    \label{fig:eucl}}
    \hfill
    \subfloat[]{\includegraphics[width=0.92\linewidth]{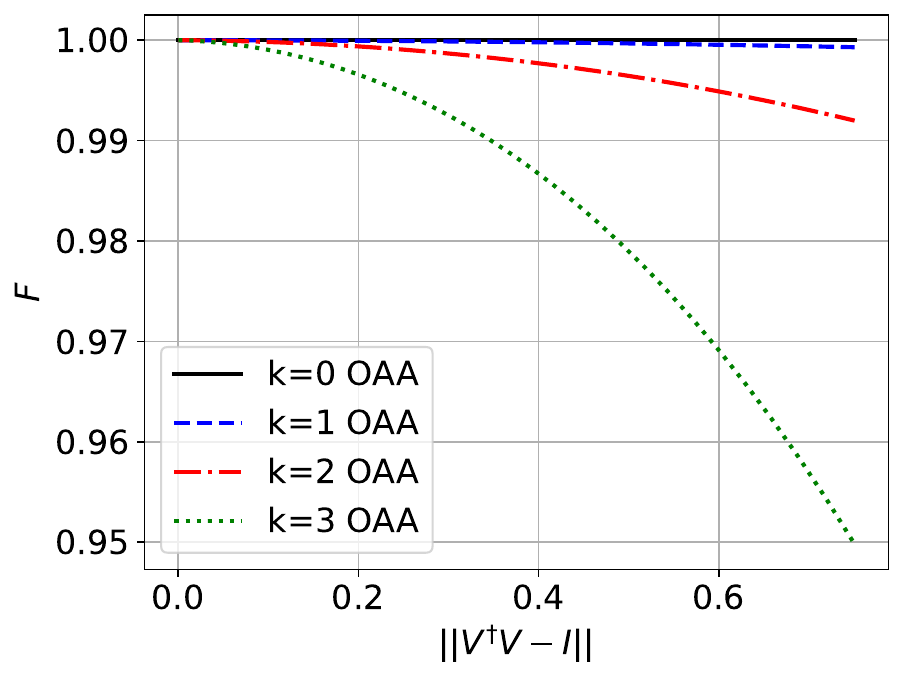}
    \label{fig:fidelity1}}
    \caption{Qiskit simulation of Euclidean distance $D$ (a) and fidelity $F$ (b) as a function of the non-unitarity parameter $\eta=\Vert V^\dagger V -\mathbb{I}\Vert$, obtained by modifying the time step $\Delta t$, for successive iterations $k$ of the OAA algorithm. Successive iterations, although required to increase success probability, also increase the error.}
    \label{fig:distance_fidelity}
\end{figure}

\begin{figure}[htbp]  
    \centering
    \includegraphics[width=0.92\linewidth]{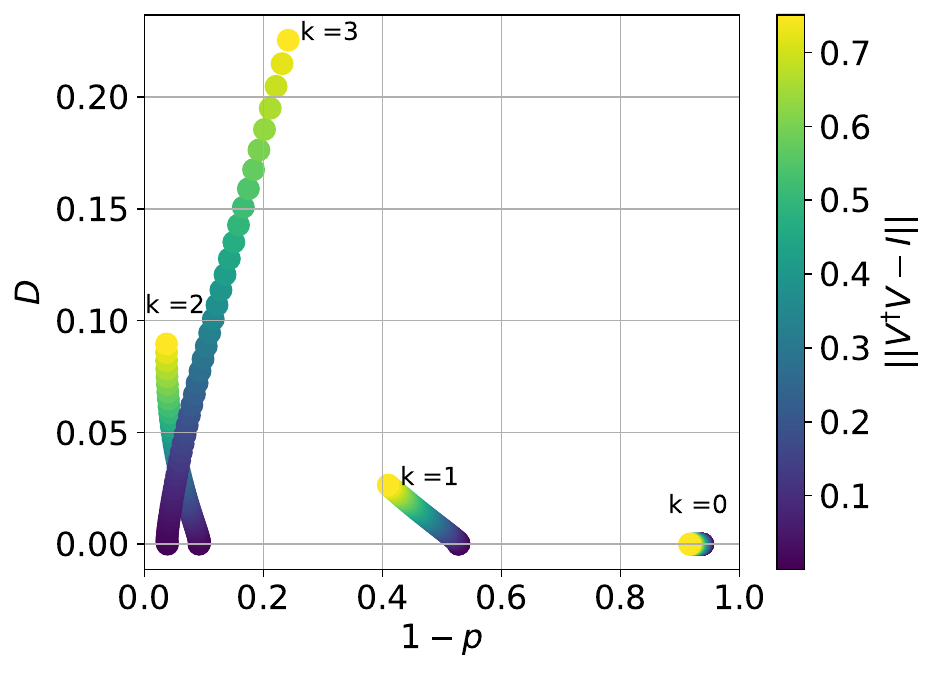}
    \caption{Qiskit simulation of the Euclidean-distance $D$ and the failure probability $1-p$ as a function of the non-unitarity of the dynamics (colour bar), obtained varying the number of OAA iterations $k$.}
    \label{fig:phase_diagram}
\end{figure}

We now focus on the error analysis due to the effect of OAA for a non unitary matrix. In Figure \ref{fig:eucl} we show the Euclidean distance~\eqref{eq:dist-func} between the exact and the approximated solution after $k$ iterations. Quantity $D$ is plotted as a function of the non-unitarity parameter $\eta$, which we remind that is zero when $V^\dagger V=\mathbb{I}$, while the number of iterations is indicated with the integer $k$, and $k=0$ indicates that no amplitude amplification is performed. At each iteration of the OAA algorithm a distortion is introduced which increases the distance of the resulting state from the exact one. 
The numerical result is in agreement with the analytical calculation in Eqs. \eqref{eq:err}  and \eqref{eq:dist-func}. 
Figure \ref{fig:fidelity1} shows the fidelity $F$ for the same system. It is evident that the decreasing function matches the analytical behaviour reported in \eqref{eq:fid-func} shown in Figure \ref{fig:func}. 
The plot suggests that Eq. \eqref{eq:proj} remains valid throughout successive amplitude amplification iterations, by substituting an increased  value for the parameter $c$.
Comparing the results for $k=1$ with the bounds on the maximum Euclidean distance and the minimum fidelity,  calculated with $\sin\theta\simeq1/4$, we find that the upper (and lower) bounds $D_{\max}$ of Eq.~\eqref{eq:dmax} (and $F_{\min}$ in Eq.~\eqref{eq:fmin}) offer conservative error estimates, as the resulting Euclidean distance and fidelity stay, respectively, significantly below and above the limit values.
The results are summarized in the probability-Euclidean distance diagram in Figure \ref{fig:phase_diagram}. The goal is to drive the state from the initial point with low success probability $p$ (and high probability of failure $1-p$ ) to a final position in the diagram with high probability, while ensuring a small error. Successive iterations $k$ of the algorithm increase the probability of success $p$ up to a value close to 1, reached when $k$ approaches the optimal number of iterations $\mathcal{O}(\sqrt8)$. The ideal error-free scenario is obtained when $1-p=0$ and $D=0$, corresponding to the origin of the plane. The effect of non-unitarity is to increase the Euclidean distance $D$ at each iteration of the OAA algorithm, thus hindering its practical use for simulating multiple time steps.

\subsection{OAA for a uniform distribution of random states}
\begin{figure}[htbp]  
    \centering
    \subfloat[]{\includegraphics[width=0.92\linewidth]{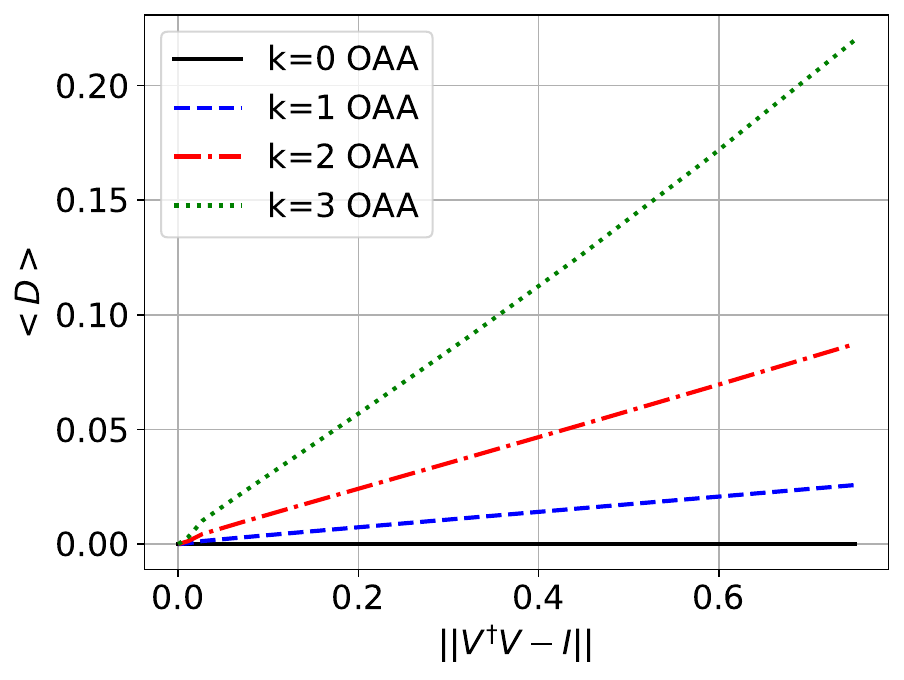}
    \label{fig:mean-dist}}
    \hfill
    \subfloat[]{\includegraphics[width=0.92\linewidth]{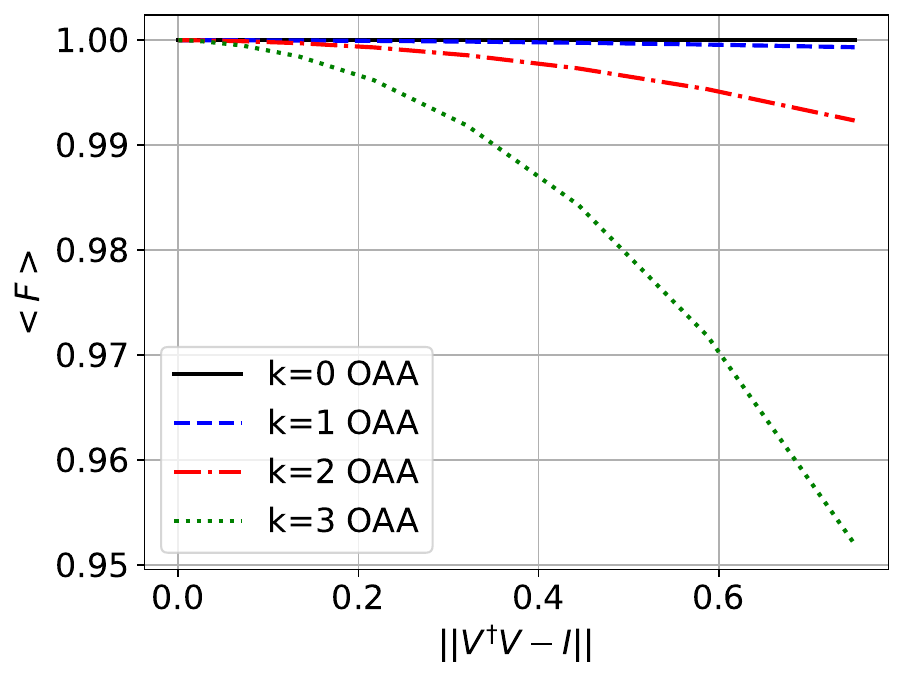}
    \label{fig:mean-fid}}

    \caption{Qiskit simulation of the mean Euclidean distance $\langle D\rangle$ (a) and of the mean fidelity $\langle F\rangle$ (b) between exact and approximate solutions, measured over $1000$ uniformly sampled random initial states, represented as a function of the non-unitarity parameter, which is changed by varying the time step $\Delta t$, for successive iterations $k$ of the OAA algorithm.}

    \label{fig:mean}
\end{figure}

\begin{figure}[H]  
    \centering
    \includegraphics[width=0.92\linewidth]{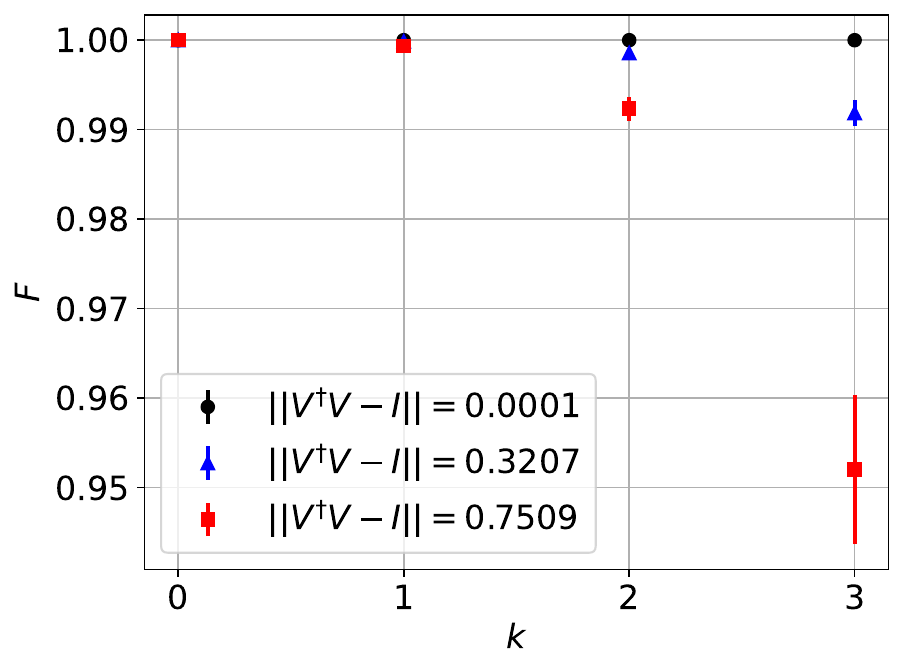}
    \caption{Qiskit simulation of fidelity average (point) and standard deviation (bar) for 3 representative values of the non-unitarity, varying $k$.}
    \label{fig:barplot}
\end{figure}

We now consider the effect of the error introduced by OAA, when applied to a non-unitary matrix, on a set of states randomly sampled from the uniform distribution on the unitary group of dimension $N=2^n$, according to the Haar measure \cite{zyczkowski2001induced}. we use $n=4$ qubits, a set of 1000 random states and the same parameters of the localized initial state case study. We point out that using a reduced number of states (e.g. 100) leads to nearly identical results, indicating that the sample size is statistically significant. 
The results are presented in Figure \ref{fig:mean}, where we show the mean values of Euclidean distance $\braket{D}$ and fidelity $\braket{F}$, obtained varying $\eta$ for a different number $k$ of iterations. The effect of successive iterations is to increase the error for all random states. Also in this case the estimate in \eqref{eq:dist-func} is valid and we retrieve the same behaviour of the fidelity in relation to the non-unitarity as for the localized case. 
The expected mean fidelity, using the expression derived in \eqref{eq:fid-func} simply coincides with the average over all possible states of the equation considering that $c=c(\psi, V)$. 
To provide a quantitative measure of the introduced error, in Figure \ref{fig:barplot} we show the average and the standard deviation of the fidelity as a function of $k$.  We observe that for increasing value of $\eta$ and successive iterations of the algorithm, the variability of the introduced error becomes larger.
Furthermore, if we use the maximum distance and minimum fidelity bounds provided in \eqref{eq:dmax} and \eqref{eq:fmin} for the specific case $\sin\theta\simeq1/4$ and compare them with our results for $k=1$, the introduced error is well controlled by our conservative estimates.

\subsection{Approximate reflection operator}
Finally, we study the effect of applying the approximate amplitude amplification algorithm, introduced in section \ref{error-mitigation}, to the case when an initial state is not fully available but a corresponding estimate may be obtained at a sufficiently low computational cost. 
\begin{figure}[htbp]  
    \centering
    \begin{quantikz}
        \lstick{$\ket{0}$} & \qwbundle{m} & 
        \gate[wires=2][0.8cm]{U}  &  \gate[wires=1][0.8cm]{-R} & \gate[wires=2][0.8cm]{U^\dagger}
         & \gate[wires=2][0.8cm]{\tilde{R}_s} & \gate[wires=2][0.8cm]{U} &\\
        \lstick{$\ket{\psi}$}& \qwbundle{n} & \gateinput[1]{}
        \gateoutput[wires=1]{}  & \gateinput[1]{} \gateoutput[wires=1]{} & & \gateinput[1]{} \gateoutput[wires=1]{} & \gateinput[1]{} \gateoutput[wires=1]{} &  \\
    \end{quantikz}
    \caption{Quantum circuit for one iteration of the amplitude amplification algorithm using an approximate reflection about the initial state $\tilde{R_s}$ applied after the initial operator $U$ that block-encodes matrix $A$.}
    \label{fig:maa-scheme}
\end{figure}
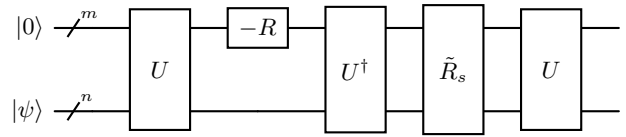
\begin{figure}[htbp]  
    \centering
    \includegraphics[width=0.92\linewidth]{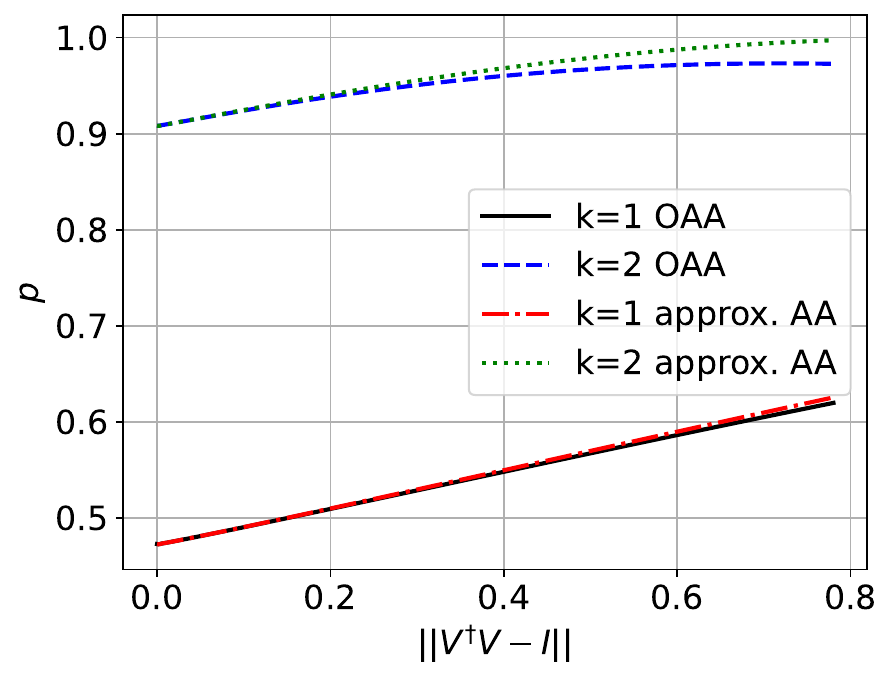}
    \caption{Qiskit simulation of success probability of measuring the $\ket{0}$ state in the ancillary qubits using two iterations of OAA and of approximate algorithm, varying the non-unitarity parameter $\eta=\Vert V^\dagger V -\mathbb{I}\Vert$ by selecting a different time step $\Delta t$.}
    \label{fig:maa-prob}
\end{figure}
\begin{figure}[htbp]  
    \subfloat[]{\includegraphics[width=0.92\linewidth]{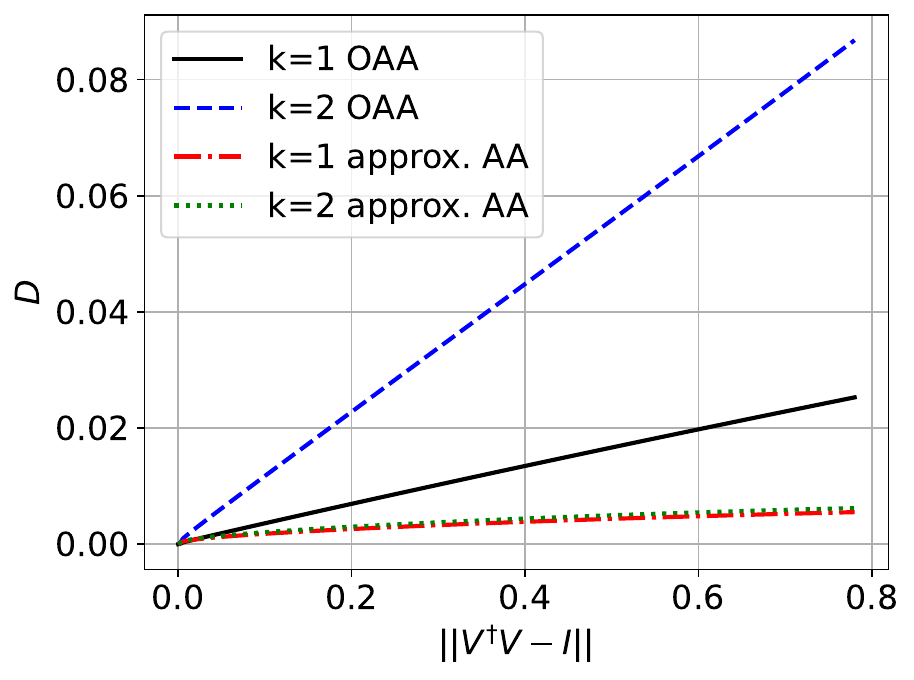}
    \label{fig:eucl-comp}}
    \hfill
    \subfloat[]{\includegraphics[width=0.92\linewidth]{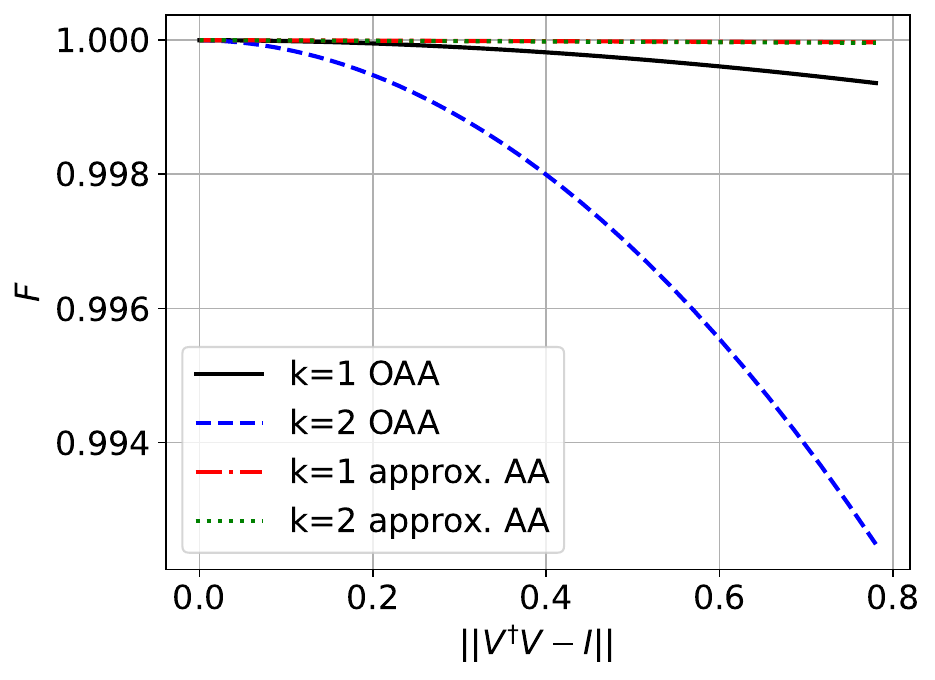}
    \label{fig:fidelity-comp}}

    \caption{Qiskit simulation of Euclidean distance $D$ (a) and fidelity $F$ (b) using two iterations of OAA and of approximate algorithm, varying the non-unitarity parameter $\eta=\Vert V^\dagger V -\mathbb{I}\Vert$ by selecting a different time step $\Delta t$.}

    \label{fig:maa}
\end{figure} 
We remind that the algorithm consists in the application of a suitable Grover operator $\tilde G$, defined in ~\eqref{eq:approx_G}, which employs a reflection $-R$ about the $\ket{0}$ state of the ancillary qubits together with an approximate initial state reflection operator $\tilde R_s$.
The circuit for implementing this approximate amplitude amplification is sketched in Figure \ref{fig:maa-scheme}.
We consider an advection-dominated scenario with parameters set to $\gamma_d=\gamma_r=0.005$ and $\gamma_a=0.9$, corresponding to a value of $\eta\simeq0.78$.  For this analysis, we employ $n=4$ working qubits and a fully localized initial state $\ket{\bm{\phi}(t_0)}$. 
As in previous cases, the non-unitarity parameter can be changed by reducing the time step $\Delta t$ keeping fixed all the other variables involved in the definition of the Courant numbers.

After the first time step, the state $\ket{\bm{\phi}(t_0+\Delta t)}$ can be obtained in the working register without any error, as the reflection $\tilde R_s=R_s$ about  $\ket{\bm{\phi}(t_0)}$ can be implemented exactly. In order to proceed with the simulation, we need to approximate $\ket{\bm{\phi}(t_0+\Delta t)}$ because the information encoded in the quantum register is not available beforehand. An easy-to-compute estimate can be achieved by considering only the effect of the advection (i.e. setting $\gamma_d=\gamma_r=0$). This choice yields an estimate $\ket{\tilde{\bm{\phi}}( t_0+\Delta t)}$, derived with minimal computational effort corresponding to a spatial translation of the  initial condition. Using this approximation, the reflection operator $\tilde{R}_s$ can be built in advance using definition ~\eqref{eq:approx_R}.
Both algorithms are able to increase the success probability as shown in Figure \ref{fig:maa-prob}.
In this setup, the approximate algorithm outperforms the OAA algorithm showing significant improvements in terms of both the Euclidean distance and the fidelity for nearly all values of non-unitarity, while still increasing the probability of success in a similar manner (see Figure \ref{fig:maa}).
Moreover, the approximate algorithm introduces an error which does not depend on the number of iterations $k$, a major improvement with respect to the OAA scheme.

\section{Conclusions}
Amplitude amplification algorithms can be used to enhance the probability of success of implementing a non-unitary matrix with the block encoding technique. However, for a majority of cases, an oblivious approach, unaware of the initial state of the quantum register is required to enhance the success probability of the quantum update. This comes with the major drawback of introducing a distortion error on the updated state at each iteration. 
For the case of the explicit Euler scheme \eqref{eq:time}, this error can alter the solution significantly, thereby inhibiting the practical use of the proposed algorithm for simulating transport problems on quantum computers with a time-step approach. To the best of our knowledge, this work provides the first quantitative estimate of the error introduced by the OAA algorithm when applied to non-unitary matrices. Moreover, we also provide additional error bounds for the Euclidean distance and fidelity of the state.
Our findings indicate that further improvements to the algorithm are inherently constrained by design, so that the development of a completely error-free OAA algorithm for non-unitary matrices remains an open problem. \\
However, it has been shown that an approximate version of the amplitude amplification, based on an estimate of the initial solution to construct an approximate reflection operator, can outperform the OAA algorithm providing a smaller distortion while still increasing the probability of success. This improvement is likely to apply to a broad class of partial differential equations related to transport phenomena. However, further improvements are required to take the success probability to the level of sustaining a viable quantum simulation of transport processes.
Our study lays the ground to the development of innovative amplitude amplification quantum algorithms, offering a new paradigm towards a viable quantum simulation of classic transport problems.

\begin{acknowledgments}
SS and CS acknowledge financial support from National Centre for HPC, Big Data and Quantum Computing (Spoke 10, CN00000013).
The authors gratefully acknowledge discussions with A. Ralli, W. A. Simons and P. Love.
\end{acknowledgments}

\bibliography{apssamp}
\onecolumngrid  

\newpage
\appendix
\section{Approximate expression for the state after one OAA iteration}\label{appendix}
Here we show how to calculate the approximate expression in \eqref{eq:appr}
Starting from 
\begin{equation}
    \ket{\Phi}=\ket{0}V\ket{\psi}
\end{equation}
\begin{equation}
     U\ket{\Psi}=\sin\theta\ket{\Phi}+\cos\theta\ket{\Phi^\perp}
\end{equation}
\begin{equation}
     U\ket{\Psi^\perp}=\cos\theta\ket{\Phi_\epsilon}-\sin\theta\ket{\Phi^\perp}=\cos\theta\ket{\Phi}-\sin\theta\ket{\Phi^\perp}-\cos\theta\ket{\epsilon}
\end{equation}
\begin{equation}
     \ket{\Psi}=\sin\theta U^\dagger \ket{\Phi}-\cos\theta U^\dagger \ket{\Phi^\perp}
\end{equation}
\begin{equation}
     \ket{\Psi^\perp}=\cos\theta U^\dagger  \ket{\Phi_\epsilon}-\sin\theta U^\dagger \ket{\Phi^\perp}=\cos\theta U^\dagger  \ket{\Phi}-\sin\theta U^\dagger \ket{\Phi^\perp}-\cos\theta U^\dagger \ket{\epsilon}
\end{equation}
and multiplying the previous two equations by $\cos\theta$ and then by $\sin\theta$, we get
\begin{equation}
     U^\dagger \ket{\Phi}=\sin\theta\ket{\Psi}+\cos\theta\ket{\Psi^\perp}+\cos^2\theta U^\dagger\ket{\epsilon}
\end{equation}
\begin{equation}
      U^\dagger\ket{\Phi^\perp}=\cos\theta\ket{\Psi}-\sin\theta\ket{\Psi^\perp}-\sin\theta\cos\theta U^\dagger\ket{\epsilon}.
\end{equation}
Now, we have all the ingredients to find how the operator $S$ acts on $U\ket{\Psi}$.
We will also use $R\ket{\Psi}=\ket{\Psi}$ and the approximation 
\begin{equation}
    R\ket{\Psi^\perp}\simeq - \ket{\Psi^\perp}
    \label{eq:aprox-perp}
\end{equation}
which is an exact result when the matrix to be block-encoded $V$ is unitary.
We now follow the steps in \cite{berry2014exponential}.
\begin{equation}
    SU\ket{\Psi}=-URU^\dagger (\sin\theta \ket{\Phi}-\cos\theta\ket{\Phi^\perp})
\end{equation}
where
\begin{gather}
    SU\ket{\Psi}=-UR (\sin\theta (\sin\theta\ket{\Psi}+\cos\theta\ket{\Psi^\perp}+\cos^2\theta U^\dagger\ket{\epsilon})
    -\cos\theta (\cos\theta\ket{\Psi}-\sin\theta\ket{\Psi^\perp}-\sin\theta\cos\theta U^\dagger\ket{\epsilon})) \\
    SU\ket{\Psi} = -UR((\sin^2\theta\ - \cos^2\theta)\ket{\Psi}+2\sin\theta\cos\theta\ket{\Psi^\perp})-2\sin\theta\cos^2\theta URU^\dagger\ket{\epsilon} \\
     SU\ket{\Psi} = -UR((-\cos(2\theta)\ket{\Psi}+\sin(2\theta)\ket{\Psi^\perp})-2\sin\theta\cos^2\theta URU^\dagger\ket{\epsilon}
\end{gather}
Using the previous approximation in \eqref{eq:aprox-perp} we get
\begin{gather}
    SU\ket{\Psi} \simeq -U((-\cos(2\theta)\ket{\Psi}-\sin(2\theta)\ket{\Psi^\perp})-2\sin\theta\cos^2\theta URU^\dagger\ket{\epsilon} \\
    SU\ket{\Psi} \simeq (\cos(2\theta)U\ket{\Psi}+\sin(2\theta)U\ket{\Psi^\perp})-2\sin\theta\cos^2\theta URU^\dagger\ket{\epsilon}
\end{gather}
and finally the result
\begin{equation}
    SU\ket{\Psi}\simeq\sin3\theta\ket{\Phi}+\cos3\theta\ket{\Phi^{\perp}}-2\sin\theta \cos^2\theta\ket{\epsilon}-2\sin\theta \cos^2\theta URU^\dagger\ket {\epsilon}.
\end{equation}
\twocolumngrid  

\end{document}